**Best Practices**

# Why every solar eclipse viewing event needs a disco ball

Valerie A. Rapson[1], Alex Pietrow[2], Robert J. Cumming[3], Jill Burns[4], Ali Cotton, Yashashree Jadhav[5], Maria D. Kazachenko[6,7,8], Casper A.L. Pietrow[9], Livia M.A. Pietrow[9], Dennis Schatz[10], Elliot Severn[11], Sundar Srinivasan[12], William Thornburgh[13], Andrea Warkentin[14], Hayley Yasui[15]

[1]State University of New York at Oneonta, [2]Leibniz-Institut für Astrophysik Potsdam, [3]Onsala Space Observatory, Department of Space, Earth and Environment, Chalmers University of Technology, [4]Glen Grove Elementary School, [5]Stevens Institute of Technology, [6]Department of Astrophysical and Planetary Sciences, University of Colorado Boulder, [7]National Solar Observatory, [8]Laboratory for Atmospheric and Space Physics, University of Colorado Boulder, [9]Leiden University, [10]National Science Teaching Association, [11]SHU Discovery Science Center and Planetarium, [12]Instituto de Radioastronomía y Astrofísica, UNAM, [13]Eastern Kentucky University, [14]Round Rock Public Library, [15]Carden School of Fresno

*Solar eclipses offer unparalleled opportunities for public engagement in astronomy. Large groups of people often gather to view eclipses, and these events require affordable and easy-to-use tools to safely observe the Sun. One unique way to observe a solar eclipse is by using a disco ball. Here, we present an analysis of the experiences of educators who used a disco ball as a solar projector during various public outreach events. Through a survey conducted shortly after the April 2024 total solar eclipse and the March 2025 partial solar eclipse, we collected data on the use, engagement, and perceived educational value of a disco ball projector from 31 individual events. The results suggest that disco balls were not only affordable and safe, but also popular and educational.*

## 1. Introduction

The world is currently experiencing multiple solar eclipses where the path of totality or maximum solar coverage passes over heavily populated areas. Following eclipses visible in India and Asia in 2019 and 2020, a partial eclipse occurred over Europe and the Middle East in 2022, the Americas experienced an annular eclipse in 2023 and a total solar eclipse on 8 April 2024. In 2026-2028, Europe, Africa, and the Middle East can look forward to several more eclipse events. Solar eclipses have inspired many educational institutions to host public eclipse viewing events, featuring a variety of hands-on activities, solar telescopes, and extensive use of solar viewing glasses. They also inspired a great deal of creativity in solar observing techniques, from building solar viewing tents (Pitts, 2023) to 2-lens safe solar viewers (Richardson, 2024).

Another unique, inexpensive, and engaging tool that can be used to observe the Sun is a disco ball, also commonly known as a mirror ball. The small square mirrors that adorn a mirror ball each have the reflective equivalence of a pinhole camera aperture (Wood, 1934; Nilsson, 1986). When sunlight strikes a mirror, it acts like a reflective pinhole, creating a small projection of the Sun. The brightness of the projected solar disc image and the broad range of distances at which it comes to a focus are both dependent on the width and area of the mirror (Cumming et al., 2024). During a solar eclipse, the dark disc of the Moon is clearly visible in these projections, making it easy to watch the progression of the eclipse over time.

Cumming et al. (2024) explored the theory behind observing the Sun with a disco ball and discussed the educational potential of these tools during solar eclipses and elevated sunspot activity. However, pinhead mirrors have a longer astronomical history. The ball mirror[1] (a pinhead mirror fixed to a ball with a collar mounting) was invented by Vivek Monteiro for observing the total solar eclipse of October 1995, and subsequently developed and used during the International Year of Astronomy 2009 (Monteiro et al., 2020). It was also used for both solar eclipses and the transit of Venus in 2012 (Navnirmiti Learning Foundation, 2024).

Before the 8 April 2024 total solar eclipse, the authors published an article in the *Bulletin of the American Astronomical Society* (Rapson et al., 2024) describing how to set up a disco ball exhibit for solar observing, and briefly shared some experiences in testing the disco ball during the 2022 and 2023 solar eclipses. Overall, the libraries, schools, and observatories that utilised the disco ball experienced great success in their outreach events. Visitors were captivated by the solar projections and particularly enjoyed seeing images of the partial eclipse on nearby walls or their clothing. Families and groups could observe the eclipse projections collectively and were often intrigued by how the mirrors on the disco ball created the projections. Our initial testing suggested that a mirror ball would make a great addition to any public solar observing event, so we wanted to spread the word and gather more data and feedback from testers in a variety of settings. Our article was widely shared among the astronomy education community, with the hope that it would inspire others to set up their own mirror ball exhibit for the 8 April total solar eclipse. The paper garnered much excitement on social media, and many people exclaimed that they would happily give the disco ball a try.

Following the April 2024 eclipse, we developed a 34-question online survey to gain a deeper understanding of how people utilised the disco ball as an outreach tool during a recent event. This survey included a variety of ranking and open-ended questions that asked for details about their event, how the disco ball was used, how people interacted with it, how beneficial the users found the tool, safety concerns, and plans for future use. Survey questions were not shared with sites prior to the eclipse, but we did ask event organisers and staff to write down anecdotes from guests and take photos to share with us later. We were most interested in the collective thoughts from event organisers on how beneficial an education tool they perceived the disco ball to be, and the ease with which it could be effectively implemented. Survey availability was promoted via social media and shared with colleagues who previously expressed interest in testing a disco ball for eclipse viewing. The survey was circulated a second time shortly after the 29 March 2025 partial solar eclipse to capture additional responses.

In this paper, we summarise the survey results and explore the pros and cons of using a disco ball as an educational and public outreach tool during solar eclipse or solar viewing events. We also provide insights on how anyone with an innovative idea for public outreach can share their project with the scientific and education community. Though previous studies (e.g., [Clark et al., 2024](#)) have explored the general education benefits of solar eclipse events for people of all ages, our study specifically focuses on the use of disco balls for solar observing.

## 2. Survey respondents

After the April 2024 eclipse, our survey was shared widely across North America via social media, email Listservs of astronomy educators, and word of mouth to various groups interested in astronomy and science education. After the March 2025 eclipse, the survey was disseminated to colleagues across Europe who were in the path of the partial eclipse. Anyone who had used a mirror ball to observe the Sun during a recent solar eclipse, or had used it to observe the Sun during any public event, was encouraged to fill out the survey. Since the initial target audience was event organisers in the United States, the survey was written in English. A total of 31 unique responses were collected.

Most responses (25) were from people who used the disco ball during the 8 April 2024, solar eclipse in the United States or Mexico. Nine respondents hosted events in the path of totality, and 16 responses came from events that observed a partial eclipse. Four respondents observed the 29 March 2025 eclipse from Sweden, and six respondents also used the disco ball either during an earlier solar eclipse or a non-eclipse event. Some sites used the disco ball at multiple events. The number of people who attended the events where the disco ball was on display ranged from just a single person displaying it at their home to groups of thousands on a college campus. There was a roughly even mix of small (<100 people) and large (>100 people) events. Survey responses came from a variety of different organisational types (e.g., museums, schools, universities, libraries, personal families, churches) and people served (i.e., pre-school children through college students, and adults). One respondent used the disco ball as part of an eclipse-viewing training session for K-12 educators hosted during a workshop sponsored by the United States' National Science Teaching Association. This undoubtedly helped spread the word about using a disco ball as an educational outreach tool and encouraged schoolteachers to try it in their classrooms.

**3. Spreading the word about innovative educational techniques**

The One of our main goals in the lead-up to the 8 April 2024 eclipse was to share this idea broadly with other people, and to emphasise its use to educators in particular. A draft of [Cumming et al. (2024)](#) was posted to arXiv.org in Fall 2023 and garnered significant press coverage from US media sites, including Space.com ([Specktor, 2023](#)) and Futurism ([Noor, 2023](#)). It was also featured in Astrobites ([Clarke, 2023](#)), an astrophysical literature journal that summarises astrophysics research in a way that is accessible to undergraduate students and science enthusiasts. The [Cumming et al. (2024)](#) paper published in *Physics Education* was highlighted by *Physics World* in January 2024 ([Hiscott, 2024](#)). Despite the media hype, our survey respondents reported that most of them had heard about the disco ball technique through colleagues or social media posts referencing either the [Cumming et al. (2024)](#) *Physics Education* paper or the [Rapson et al. (2024)](#) article in the *Bulletin of the American Astronomical Society*. This was surprising, as we thought that news articles would have played a larger role in promoting the use of disco balls. It's possible, though, that we are seeing some bias in this result. Our survey was shared extensively via social media to gather responses. Those who saw the survey on social media may have been more likely to see the disco ball information via social media, as well. One should not discount the power of social media in spreading new and innovative educational ideas, especially for schoolteachers and informal educators (e.g., [Duprow, 2024](#); [Mercado & Shin, 2025](#), and references therein).

Another key component that allowed our idea to spread is that the concept of solar projections off mirrors, similar to a disco ball, is very easy to test and understand. Mirror reflections are something we experience every day, and the idea that a mirror could project an image of the Sun was intuitive for most people. While the exact optical principles that make this possible are complex ([Cumming et al. 2024](#)), one does not need to understand pinhole optics to try or enjoy the experiment's results. This likely made the technique more accessible to educators and astronomy enthusiasts without a need for deep understanding of optics. Thus, people were more willing to incorporate it into their activities. In fact, one survey respondent who showcased the disco ball during a teacher training session prior to the April 2024 eclipse stated that, "Participating educators saw photos of the images produced by the disco ball. All were amazed at what they saw, and many immediately went online to buy a disco ball." While not every innovative educational tool is going to be simple to test, they should be presented in such a way that educators with a basic understanding of science can easily set up the experiment and explain the results to the general public. Having more details available in scientific publications for those who are interested in learning more is an added benefit.

## 4. Mechanics of using the disco ball

In our survey (found in Appendix A), we asked respondents to share details about their disco ball setups, such as the number of mirror balls on display, how much money they spent on the exhibit, how and where they were displayed, and what other tools were used to observe the Sun at their event. Most sites (80.6%) only had one disco ball on display, regardless of the anticipated number of attendees. The disco balls were set up in a variety of ways: some were hung freely outdoors, others were placed on stationary podiums, and some groups had staff or the visitors themselves walk around with the disco balls in hand (Figure 1). Some respondents positioned their disco ball near a window so that they could easily view the projections on a wall indoors without having to be outside during the event. Another group hung the disco ball inside the dome of their observatory, causing the projections of the Sun to fill the dome ceiling (Figure 2). This allowed visitors to observe the Sun through a filtered optical telescope while being surrounded by projections of the solar eclipse. Another organisation hung small disco balls from the awnings of local shops and informed shopkeepers that the disco balls would become a fun teaching tool on eclipse day. Regardless of the type of setup, the disco balls were frequently used in conjunction with other observing tools, like solar glasses, pinhole projectors, or telescopes.

The disco ball displays, in general, were very low-cost. Of survey respondents, 74.2% said they spent less than $25 USD, and only two spent more than $100 USD. The two organisations that spent the most money either purchased multiple disco balls for their display or had grant funding for an extensive exhibit. For example, one University received grant funding to build an "eclipse experience marquee" that included pinhole projections of the Sun under a tent, and projections from a larger glass disco ball. Respondents shared that small, party-style disco balls were just as effective as larger professional ones – providing they had glass mirrors – and were easier to obtain from online retailers. In the United States, disco ball-themed fashion and décor happened to be very popular around the time of the April 2024 eclipse. This was likely due to celebrities and influencers pushing the trend of wearing or displaying reflective materials that created disco ball-like effects (e.g., Nolfi, 2023), and probably had nothing to do with their scientific usefulness during the eclipse. Nevertheless, it was easy to find decorative and inexpensive flower pots and candles covered in mirrors, as well as small disco balls, at large retail stores. This may have made it easier for sites to acquire the necessary supplies to set up an exhibit at the last minute, and increased the number of sites that had a disco ball item on display.

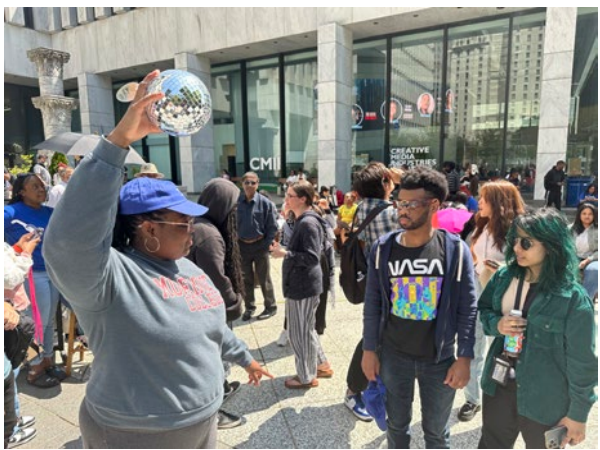

**Figure 1** *An educator walking around an eclipse event showing off the disco ball projections. Credit: Misty Benz*

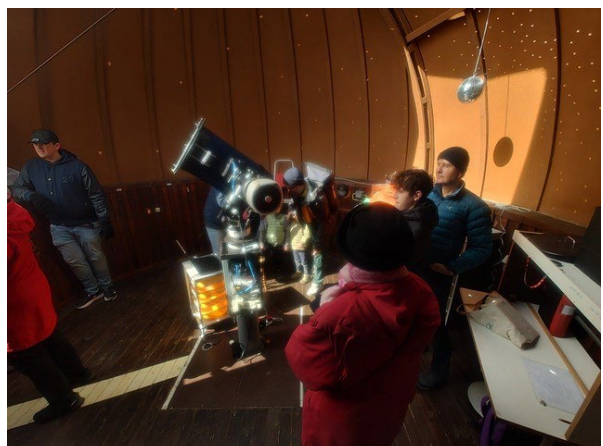

**Figure 2** *Solar eclipse projections on the inside of an observatory dome created by a hanging disco ball. Credit: Göran Kajler*

**5. Disco ball promotion**

To determine if the disco ball, in particular, drew people to an observing event, we asked respondents if they had promoted its use in advance. Only 25.8% of people advertised the use of a disco ball at their event, and most just used word of mouth with visitors whom they knew would be attending the event anyway. One university promoted the disco ball on event posters and shared this specific research project in media interviews. This resulted in a few adults stating that they were particularly excited to see the disco ball projections, but they did not come exclusively because of the disco ball project. Another organisation specifically limited the amount of disco ball advertising for fear that it would draw in a crowd beyond their physical capacity. Most other organisations that had disco ball promotions were schools, where teachers informed the classes of students ahead of time that one of the tools they could use to observe the eclipse would be a disco ball. From our results, advertising the use of a disco ball does not necessarily increase event attendance, but it can promote additional excitement amongst children and adults who plan to attend a solar eclipse viewing event.

**6. Disco ball engagement and visitor experience**

To better understand how many people interacted with the disco ball and assess its value as an educational tool, we asked survey respondents to estimate the percentage of people they believe engaged with the disco ball, provide open-ended comments on its effectiveness, and share their opinion on whether it was worthwhile to display. Figure 3 shows a histogram of the percentage of visitors who engaged with the disco ball. Nearly half (41.9%) of respondents said that a majority (75-100%) of their visitors interacted with the disco ball display, while others reported lower engagement numbers. The percentage of visitors who engaged with the disco ball seems to be most strongly correlated with the total number of visitors at the event (Figure 4). All groups with < 10 visitors report 75-100% engagement, and larger groups > 500 tended to report less than 75% engagement. It is likely that for large events, one disco ball is insufficient to attract or engage all visitors, and we cannot be certain that all visitors will even notice the disco ball exhibit. Small groups tended to feature the disco ball prominently and shared that they often took the time to explain the purpose of the disco ball to the group and let visitors hold it. Thus, the disco ball would be an excellent tool to use in a small classroom setting where everyone has a chance to see and engage with the projections.

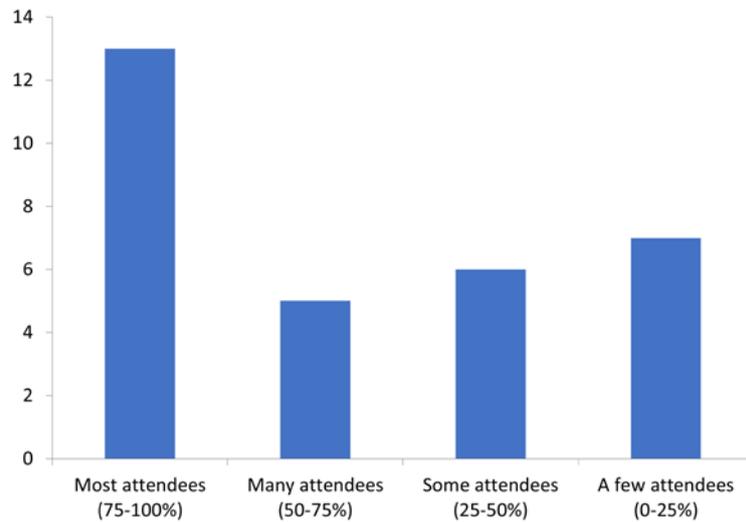

*Figure 3* Survey results for the question "Approximately how many attendees do you think interacted with the disco ball display?"

The small pinhead-like mirrors require a strong light source to make an obvious projection. Two sites reported low engagement levels, citing mostly cloudy skies that hindered visitors' ability to observe projections effectively. In instances of partly cloudy skies, it is more beneficial to use solar glasses or white-light telescopes where visitors can still see the Sun, even through thin clouds. Other occurrences of lower engagement were attributed to poor planning for the location and display of the disco ball. Many sites reported that they wish they had tested the projections ahead of time to ensure they would fall within a slightly darkened corridor, making them more visible to visitors. It is possible that many people noticed the disco ball at these events, but may not have seen the projections clearly. In this case, the disco ball enhances the party atmosphere of an event, but does not provide any educational value.

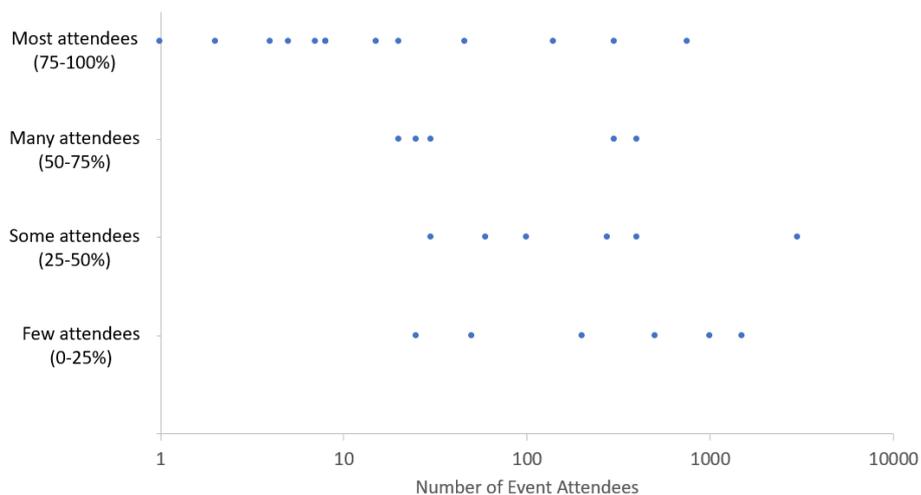

*Figure 4* Correlation between the percentage of people who interacted with the disco ball projections and the number of attendees at an event.

It is interesting to note that, when asked about all the tools used to observe the eclipse, 5 of our 31 respondents did not select any additional observing tools on our survey; two of whom specifically stated that they chose to use no other tools. All five of these respondents were groups of <10 people and indicated that they had a positive experience during the eclipse. This suggests that projections using a disco ball are viewed as equally exciting as other methods of pinhole projection and may be easier to create and share with small groups.

We additionally asked people to reflect on whether or not people understood that the disco ball was creating visible projections of the Sun and the solar eclipse. Only 29.0% said that "most visitors recognized on their own that the disco ball was projecting a solar image", whereas 51.6% of respondents said that visitors "recognized the solar projections after reading posters or interacting with a staff member who explained what was happening". In a related open-response section, one respondent stated that "With a clear sign explaining how to use it, one can increase the number of people using [the disco ball]." This need for additional explanation is typical of science museum exhibits. Visitors often figure out how to engage in a hands-on exhibit, and then look towards signage or a docent for an explanation of the science behind the exhibit (e.g., Weiss, 2013). Again, sites with less-than-ideal weather reported that visitors required an explanation as to the disco ball's purpose, especially when the projections were not prominently visible. There was also no apparent correlation between age or type of visitors and whether or not they were able to recognise the eclipse projections on their own. Some groups of young children immediately recognised the crescent shape of the Moon covering the Sun, and some groups of adults required an explanation to make the connection. Events at college campuses or institutes of higher learning, where students were the primary attendees, also reported a mix of literacy when it came to interpreting the solar projections. Slater & Gelderman (2017) point out that students of all ages often have misconceptions about how and why eclipses occur, and thus may have difficulty understanding images or projections of an eclipse without additional explanation. Interestingly, none of our respondents reported seeing common misconceptions, such as mistaking the projected solar image for the Moon. The relatable idea that the Sun is the bright object reflecting off the disco ball mirrors creating the images, and the Moon was the dark disc appearing to block the Sun, may have made the projections easier to interpret.

Respondents provided some excellent feedback regarding how their visitors engaged with the disco ball. One university stated:

> "The children loved the disco ball and played with it for hours before the event… It [was] a great addition to the visual storytelling for the eclipse, as well as an interactive display."

A research institution that hosted an event shared that:

> "The disco ball was a very novel idea, and it was enjoyed by all viewers. The large number of images projected onto indoor and outdoor walls enabled people to view these images without overcrowding."

A University that held an event specifically for secondary school children said that:

> "The aesthetic effect is really nice, especially in our big exhibition hall. Most people were not particularly curious about how it worked, I think because it was so obvious."

And one respondent simply stated

> "It made everyone happy."

Common themes of accessibility, safety, and ease of viewing with a large group were also prevalent in the comments. Compared to observing through solar glasses or a telescope, people generally liked the community aspect of the mirror ball projections. Eclipse glasses and solar filters block out everything except the Sun's bright disc, making it an isolating experience. With a disco ball, families and children could all look at the eclipse together and have discussions about what they were seeing and why. It also encouraged people to take pictures of the projections on the ground, or especially when they fell on the clothing of another person (Figure 5). For sites with mobile disco ball displays, children had a very enjoyable time carrying the disco ball around and explaining the projections to other visitors. This empowered visitors to take control of their experience and, in some instances, become the educators themselves.

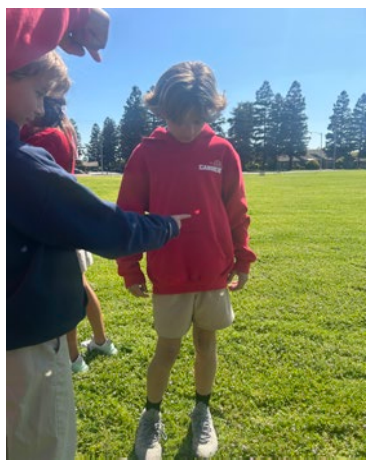

**Figure 5** *A group of students observing a projection of a solar eclipse on a child's shirt. Credit: Hayley Yasui*

An interesting anecdote about the mirror ball projections came from a respondent who travelled with a group to Milwaukee, Wisconsin, for the 8 April 2024 solar eclipse. They pointed out that during prior solar eclipses in the US (namely, the 21 August 2017 eclipse), they recalled seeing many projections of the eclipse on the ground underneath trees. This phenomenon is caused by small gaps between the leaves on the trees acting as natural pinhole projectors. But in April, most of the trees in Wisconsin had not sprouted leaves yet, so there were no natural eclipse projections around. The disco ball, however, provided essentially the same type of projection effect and allowed their group of approximately 200 people to enjoy the solar eclipse together. Though the seasonal differences between eclipse viewing techniques were not something we initially considered, the realisation is an important one. For solar eclipses that occur in winter and in locations where trees are bare or nonexistent, a mirror ball can serve as a great substitute for the natural pinhole projections of the Sun that the trees would otherwise create.

When asked if they felt having a disco ball display was worthwhile, 90.2% of respondents said it was "Extremely" or "Somewhat worthwhile" (71% and 19.4%, respectively; Figure 6). The two respondents who provided a neutral answer to this question were from sites that experienced poor weather, so the disco ball was not as effective as it could have been. Generally speaking, visitors and event organisers found that the disco ball display was very exciting, a positive educational tool, and highly beneficial to have.

**7. Comparison with other solar observing tools**

As mentioned above, most survey respondents used a variety of observing tools at their events (Figure 7). Solar glasses and pinhole projectors were the most commonly used tools, with a majority of respondents distributing solar glasses to visitors and half having pinhole projectors on display. Sites that had access to safe solar telescopes (35%) also displayed them, and 38.7% offered other hands-on science activities. Sites that anticipated larger crowds often had a larger variety of observing tools and activities available so that all visitors would have the opportunity to observe the Sun with at least one viewing method.

Prior to the recent solar eclipses, the news media heavily emphasised safe solar observing techniques and focused heavily on the use of solar eclipse glasses. It is estimated that over 100 million pairs of eclipse glasses were sold in the US in the few months before the 8 April eclipse (Sarkis, 2024). Thus, people were expecting to use eclipse glasses at these eclipse events, and some likely expected to be able to observe through solar telescopes or binoculars. But many were not expecting to see disco balls used for solar projection. Survey respondents shared that many of their visitors said something like "I never knew a disco ball could be used for science" after interacting with the exhibit. Others simply said, "This is so neat!", or "Oh my gosh, that's crazy!". Disco ball projections appear to have added a "wow" factor to many events and served as a welcome addition to other observational tools (Figure 8). They even allowed some guests to remain entirely indoors and still enjoy the solar eclipse projected through a window. For guests with mobility issues or other special needs that may not allow them to be outside during a solar eclipse, a disco ball projection aimed at an indoor wall may be the perfect way to include them in the viewing experience.

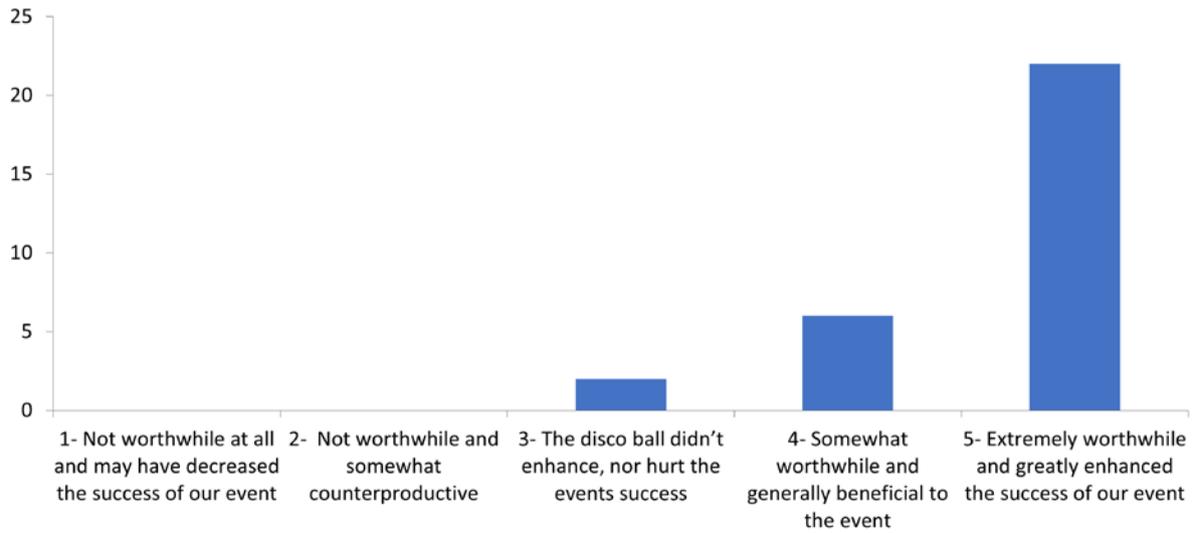

*Figure 6* Survey responses to the question "On a scale of 1-5, do you feel it was worthwhile to have a disco ball display?"

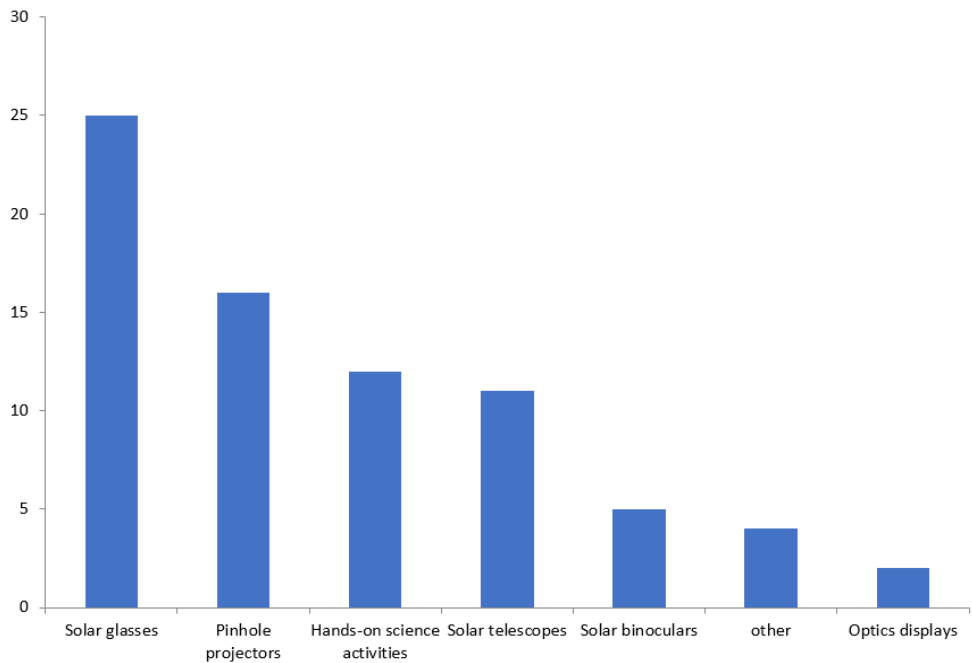

*Figure 7* Survey responses to the question "What other activities or solar viewing tools did you use at your event?" Respondents could select more than one answer.

Despite all the positive benefits of using a disco ball to observe the Sun, some organisations reported that solar telescopes and glasses were still the most popular. There is something special about seeing the Sun up close through a telescope, or seeing it with your own eyes

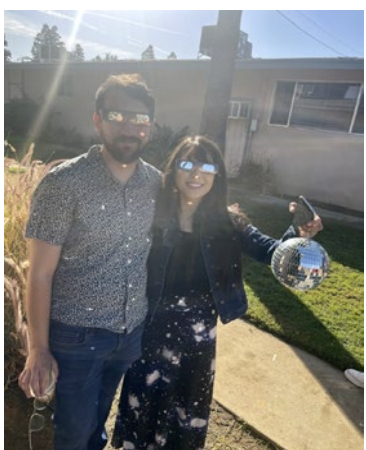

**Figure 8** *Two event visitors posing with solar glasses and a disco ball. Projected images of the eclipse adorn their clothing. Credit: Hayley Yasui*

through glasses, that a projection cannot match. For future eclipse events, we suggest having at least one disco ball on display, but also offering visitors other methods of observing the Sun as well. This will provide a variety of different experiences and allow for both educational group discussions and personal introspective experiences when observing.

**8. Safety concerns**

Safety is always the primary concern when observing a solar eclipse. Large events always pose a risk from a crowd-control perspective, and there are serious known risks of permanent eye damage from staring directly at the Sun with no protection (e.g., Chou, 2016; WLS-TV, 2016). Solar observing tools – such as mirror ball projections – that pose a low safety risk and can be used in both large and small group settings are, therefore, ideal.

Projected images are completely safe to observe, and the reflected light from the disco ball is at such a low concentration that crossing the beam path or briefly staring into it should not cause any physical damage ([Cumming et al., 2024](#)). The disco ball also creates multiple projections that people tend to gravitate towards, causing crowds to naturally disperse away from the disco ball. When asked how safe they felt the disco ball display was, 20% of survey respondents said they used the disco balls freely and considered them "completely safe", while 9 respondents said that they felt the disco ball was safer than other solar observing methods, but still had a staff member monitoring nearby. Two sites did not answer the question, but notably, no one stated that they felt the disco balls were unsafe in any way. The only danger created by a disco ball during the period of this survey was when one fell and broke at one of the largest observing events, posing a shattered glass hazard.

A couple of our respondents who were observing alongside schoolchildren shared that parents and school administrators were extremely concerned about liability. Even after the educators shared relevant literature ([Cumming et al., 2024](#); [Rapson et. al., 2024](#)) with the administration, the students were still not allowed to interact directly with the disco ball. However, they were able to observe the projections indoors, which made for a positive teaching experience. One respondent, who was working with pre-school-aged children, shared that "The light from the disco ball gave an exciting, magical atmosphere to the classroom, and discouraged children from wanting to go outside to look at the sun."

Especially when there is a concern that young children will improperly use solar glasses, the disco ball can be a great alternative to experience the progression of a solar eclipse. The students still feel like they are a part of the observing experience, even if they are not looking directly at the Sun.

### 9. Disadvantages of using the disco ball and other unintended consequences

Disco ball users had an overwhelmingly positive experience at their events and had very little to share when it came to the cons of having a mirror ball on display. Some respondents noted that the projections can be challenging to see if they fall into well-lit areas, especially if the focus distance for the image is tens of metres away. Having a darkened projection area would be beneficial and make it easier for people to observe the images. This also provides a good opportunity to include signs to educate the public about the disco ball near where the projections fall. Keeping the disco ball stationary was also a suggestion of some respondents. While a spinning disco ball looks cool, it can be hard for people to see and interpret the moving projections.

One respondent pointed out that they hung their disco ball outside their home, relatively close to their birdfeeder. The disco ball ended up being "a distraction for birds, who avoided our feeders until we took it down. Since the reaction of wildlife was one of the key experiences we were looking forward to seeing, I wonder if that was diminished by the presence of the disco ball." It is very possible that the disco ball reflections deterred certain bird species from coming nearby, and may have made their reactions to the eclipse harder to distinguish from reactions caused by the presence of a disco ball. Hanging a disco ball may also interfere with other animals in nature and hinder one's ability to accurately participate in citizen science projects, such as NASA's Eclipse Soundscapes Project (ARISA Lab, 2024). Notably, disco balls do not appear to pose a true threat to any animals, but they may disturb their natural habitat while in use.

It is important to highlight that people who suffer from photophobia or light sensitivity may be bothered by the presence of a disco ball. Bright reflections off the mirrors can aggravate this condition, making looking in the direction of the disco ball very uncomfortable. Event coordinators should be aware that some visitors may have such sensitivities and ensure that the disco ball is not so centrally located at the event that it cannot be avoided.

**10. Tips for future users and other ways to use the disco ball**

Survey respondents provided valuable feedback for improving the experience. Most respondents wished they had tested their display ahead of time to make sure that the Sun would produce bright, easy-to-see projections on a nearby surface. All groups were successful in seeing something, but, in retrospect, many said that they would have moved their exhibit to a slightly different location so that the projections would have been clearer. Additionally, it was suggested to hang the disco ball in advance, particularly if the event is held on a college campus or in a downtown area frequented by the same individuals. Seeing the regular solar projections prior to an eclipse draws attention to the disco ball, providing a comparison between the circular projections you usually see and the crescent-shaped projections on eclipse day.

Respondents also emphasised that the use of a disco ball should be promoted more heavily in the education sector and used to observe future solar eclipses. On the theme of reuse, we asked respondents if they knew they could observe sunspots using the mirror ball projections and whether or not they planned to do so. Half of the respondents said they had heard about observing sunspots via projection, while the other half had not. However, 71% said they planned to try it in the future. Those that did not plan to use their disco ball to observe sunspots in the future were organisations like libraries that were solely focused on hosting an outreach event for the recent solar eclipse, but do not otherwise do astronomy-related activities. We also asked respondents if they planned to repurpose the disco ball, either for a solar-observing-related activity or a non-solar-related exhibit; 80.5% said they would reuse the disco ball in some way, while the rest were undecided. Indeed, multiple survey respondents stated that they have already used their disco ball to observe more than one recent solar eclipse.

As an easy-to-use and inexpensive tool, the disco ball could be very versatile in helping remote and under-resourced regions observe solar eclipses. Partnerships with organisations that promote global campaigns to observe astronomical events might help bring accessible solar observing to a large number of people. Similar to recycling events for eclipse glasses, disco ball recycling or sharing could become a beneficial endeavour. We plan to reach out to these organisations to discuss possible partnerships and welcome any other ideas from the astronomical community.

## 11. Conclusion

Disco balls are inexpensive, safe and versatile tools that can be used to observe the Sun any day, and are particularly useful to project images of the Sun during a solar eclipse. Overall, having a disco ball on display during a solar eclipse or solar viewing event was found to be a fun, safe and effective way of observing the Sun. The general public was captivated and intrigued by the use of a mirror ball for science, and actively engaged with the tool at outreach events. A disco ball was particularly effective at allowing large groups of people to observe a solar eclipse communally and was a positive addition to other solar observing tools, like glasses and solar telescopes. However, we found that the disco ball was not very effective in locations that saw partly to mostly cloudy skies, and it may have interfered with other citizen science events, like observing the actions of birds and other animals during the eclipse. Testing the projections ahead of time on a sunny day is crucial for deciding on the most effective set-up location to maximise engagement.

Experiences with innovative outreach tools, like a disco ball, can and should be shared widely through social media and listservs of astronomy educators. Conducting demonstrations as part of teacher professional development events and promoting ideas to the national media are also effective ways of spreading the word. We hope that any organisation hosting public solar observing events will consider setting up a mirror ball display, and that countries worldwide will utilise a disco ball to share upcoming solar eclipses with as many people as possible.

*Acknowledgements* *The authors would like to thank all organisations that tested the disco ball exhibit idea and took the time to complete our survey and share their experiences. Thanks to Brandi Blue (Wine Country Zoological) for providing feedback about bird interactions with the disco ball. AP is supported by the Deutsche Forschungsgemeinschaft, DFG project number PI 2102/1-1. MDK thanks support from the NSF CAREER SPVKK1RC2MZ3 award.*

**Notes**

**1.** The ball mirror invented by Vivek Monteiro is a slightly different design from a disco ball, as described in the text. In all other instances in this text, we refer only to the mirror ball or disco ball.

## Appendix

**Appendix A: Solar observing using a disco ball survey**

*Your name*

*Email*

*Job title*

*Organization name*

*Type of organization (e.g., K-12 school, library, college/university, museum, etc.)*

*Location (City, State/province/territory, Country)*

*Do we have your permission to link your responses with the organization name in statements we make in the paper? (e.g. "The Science Discovery Center experienced…"). We are happy to share a draft of the paper with any survey respondents before publication.*
    ○ Yes, you may link the experiences described below with the name of the organization.
    ○ No, I'd prefer my data to be kept anonymous and used for general statements only.

*We plan to submit our paper to the Communicating Astronomy with the Public Journal (CAPjournal). Are you interested in being a co-author on the paper, or would you just prefer an acknowledgement? Checking "Yes" to be a co-author means you agree to read the draft manuscript and provide feedback and edits before and after submission to the journal.*
    ○ Yes, I'd like to be a co-author and agree to provide feedback on the manuscript.
    ○ No, I do not want to be a co-author and am happy with an acknowledgement at the end of the paper.

*How did you first hear about using a disco ball to observe the sun?*
    ○ Read the "Why every observatory needs a disco ball" paper on ArXiV
    ○ Read the "Why every observatory needs a disco ball" paper in Physics Education
    ○ Read the "Solar Disco" paper in the Bulletin of the American Astronomical Society's eclipse edition
    ○ Online news article (e.g. space.com, phys.org)
    ○ Social media post from a community group or online news website
    ○ Colleague/friend shared the paper or a news article with me
    ○ I noticed the effect myself

*Please select the event(s) in which you used a disco ball to observe a solar eclipse. (Select all that apply)*
- ☐ Solar eclipse of October 25, 2022 (I was in Totality)
- ☐ Solar eclipse of October 25, 2022 (I saw a partial eclipse)
- ☐ Solar eclipse of October 14, 2023 (I saw the full annular eclipse)
- ☐ Solar eclipse of October 14, 2023 (I saw a partial eclipse)
- ☐ Solar eclipse of April 8, 2024 (I was in Totality)
- ☐ Solar eclipse of April 8, 2024 (I saw a partial eclipse)
- ☐ Partial solar eclipse of March 29, 2025
- ☐ I used the disco ball at a solar observing event not related to a solar eclipse

*Approximately how many attendees were at your event?*

*Who was the main audience at your event? (Select all that apply)*
- ☐ *Elementary School children (Ages 5-11)*
- ☐ *Secondary School children (Ages 12-17)*
- ☐ *Families/general public*
- ☐ *College students*
- ☐ *Adults*
- ☐ *Other*

*Approximately how many attendees do you think interacted with the disco ball display?*
- ○ Most attendees (75-100%)
- ○ Many attendees (50-75%)
- ○ Some attendees (25-50%)
- ○ A few attendees (0-25%)

*Where did you set up the disco ball exhibit?*
- ○ Indoors
- ○ Outdoors
- ○ Indoors and Outdoors

*How many disco balls did you have on display?*
- ○ 1
- ○ 2-3
- ○ >3

*Please approximate the total amount of money spent on your disco ball exhibit. Include the cost to purchase the disco balls, signage, and other tangible items related to the display. Do not include staff costs.*

*Describe the types of disco balls used (large mirror/small mirror, round balls, other shapes) and the set-up for your disco ball exhibit (specific location, signage, etc.)*

*How were the disco balls displayed or mounted during your event? (Select all that apply)*
- ☐ Free hanging (e.g. ball hung from the ceiling)
- ☐ Stationary display (e.g. ball mounted on a pedestal)
- ☐ Mobile (e.g. staff member walked around with the disco ball)
- ☐ Other

*Was the use of a disco ball at your event promoted ahead of time?*
- ○ Yes
- ○ No

*If the disco ball display was promoted ahead of your event, in what ways was it promoted (posters, social media, etc.)? Do you think this increased attendance at your event?*

*How did visitors engage with the disco ball display? Did they enjoy it? Please be as detailed as possible.*

*Did visitors understand the purpose of the disco ball as a solar projection device? Or were staff members necessary to explain the concept?*
- ○ Most visitors recognized on their own that the disco ball was projecting a solar image.
- ○ Most visitors recognized the solar projections after reading posters or interacting with a staff member
- ○ Few visitors recognized that the disco ball was projecting a solar image, and did not further engage with the exhibit
- ○ Do you have any specific quotes or anecdotes from visitors related to your disco ball display that you would like to share?

*Please describe your overall experience using the disco ball as a tool to observe the eclipse and educate the general public.*

*On a scale of 1-5, do you feel it was worthwhile to have a disco ball display?*
- ○ 1- Not worthwhile at all and may have decreased the success of our event
- ○ 2- Not worthwhile and somewhat counterproductive
- ○ 3- The disco ball didn't enhance, nor hurt the events success
- ○ 4- Somewhat worthwhile and generally beneficial to the event
- ○ 5- Extremely worthwhile and greatly enhanced the success of our event

*What other activities or solar viewing tools did you use at your event? (Select all that apply.)*
  ☐ Solar glasses
  ☐ Solar binoculars
  ☐ Solar telescopes
  ☐ Pinhole projectors
  ☐ Hands-on science activities
  ☐ Optics displays
  ☐ Other

*How effective did you feel the disco ball projections were compared to other tools used to observe the sun? Is there anything specific you'd like to share about how visitors interacted with the disco ball vs. other tools?*

*Safety is a big concern for any event that involves observing the sun. How did you ensure safe viewing of the sun during your event? (Select all that apply.)*
  ☐ Provided eclipse glasses to all visitors
  ☐ Did not allow visitors to bring their own solar viewing equipment
  ☐ Utilized staff to monitor all solar viewing stations
  ☐ Used a mixture of projection devices and direct observing tools
  ☐ Only used projection devices to observe the sun (no direct solar viewing)
  ☐ Limited the age of visitors to your event (i.e. imposed an age cut-off for children)
  ☐ Other

*What role did solar projections from a disco ball play in your safety plan?*
  ○ We didn't use disco balls because of safety concerns
  ○ We closely monitored disco balls because of safety concerns
  ○ We felt the disco balls were safer than other methods, but still had a staff member monitoring nearby
  ○ We used disco balls freely with little monitoring because we considered them completely safe

*Were there any cons to using the disco ball to observe the sun? Is there anything you would change about your disco ball exhibit if you were to run an event like this again?*

*Do you have any advice for other sites that will experience a solar eclipse in the near future and may want to have a disco ball display?*

*Did you know that you can observe sunspots in solar projections from a disco ball?*
  ○ Yes, and I plan to attempt to do that in the future
  ○ Yes, but I do not plan to use the disco ball for that purpose in the future
  ○ No, but I plan to try observing sunspots with the disco ball in the future

○ No, and I do not plan to try to observe sunspots with the disco ball in the future

*Will you be repurposing the disco balls for another type of event/exhibit now that the eclipse is complete?*
○ Yes, we will use the disco balls for future solar observing events or for an optics display/activity
○ Yes, we will use the disco balls for another event/exhibit not related to solar observing
○ We have not yet decided if the disco balls will be repurposed
○ No, we will not be using the disco balls again. We donated or disposed of the disco balls

*Is there anything else you would like to share with us related to solar observing with a disco ball?*